\documentclass[twoside,reqno]{mbook}
\usepackage{epsfig,cite}
\usepackage{amssymb,amsmath}
\usepackage{times}
\begin{document}

\sloppy \raggedbottom

 \setcounter{page}{1}

\newpage
\setcounter{figure}{0}
\setcounter{equation}{0}
\setcounter{footnote}{0}
\setcounter{table}{0}
\setcounter{section}{0}


% use the thanks command within \title, \author or \address for footnotes;
% \title{Title} or  \title{Title\thanks{...}}
% \author{Name1}{aff.label1}, \coauthor{Name2}{aff.label2},  \coauthor{Name3}{aff.label3}
% \address{Address1}{aff.label1}
% \address{Address2}{aff.label2}
% \address{Address3}{aff.label3}
%\runningheads{Names of Authors in Initial Capitals}{Short Title in Initial Capitals}

\title{Isospin mixing, Fermi transitions and signatures of nuclear deformation within a mean field approach}
%\thanks{Support from the U.S.
%National Science Foundation, PHY-0140300, and the Southeastern
%Universities Research Association is gratefully acknowledged.}}

\runningheads{Isospin mixing, Fermi transitions and signatures of nuclear deformation...}
{R.~Alvarez-Rodriguez et al.}

\begin{start}
\author{R.~Alvarez-Rodriguez}{1}\thanks{Present address: Institut for Fysik og Astronomi, DK-8000 Aarhus C, Denmark}, 
\coauthor{E.~Moya de Guerra}{1,2},
\coauthor{P.~Sarri\-guren\-}{1}, \coauthor{O.~Moreno}{1}

\address{Instituto de Estructura de la Materia,\\
Consejo Superior de Investigaciones Cient\'{\i}ficas, Madrid E-28006, Spain}{1}

\address{Departamento de F\'{\i}sica At\'omica, Molecular y Nuclear,\\
Universidad Complutense de Madrid, Madrid E-28040, Spain}{2}

%\address{Department of Physics, Rochester Institute of Technology}{3}

\begin{Abstract}
Gamow-Teller and Fermi transitions are considered in a HF+BCS+pnQRPA
theoretical framework. We show here how Gamow-Teller strength distributions
can be used in a search for signatures of nuclear deformation in 
neutron-deficient Pb, Hg and Po isotopes, as well as how Fermi 
transitions allow us to quantify isospin 
mixing in some even Kr isotopes around $N=Z$.
\end{Abstract}
\end{start}

%\section[]{Introduction}

\section{Theoretical framework}\label{frame}

Microscopic models describe the structure of the nucleus in terms
of the degrees of freedom of its microscopic constituents: the
nucleons. Our starting point is a non-relativistic Hamiltonian
containing one- and two-body interactions, 
\begin{equation}
\hat H = \sum_{ij} t_{ij} \hat a_i \hat a_j + \frac{1}{2} \sum_{ijkl}
v_{ijkl} \hat a_i^\dag \hat a_j^\dag \hat a_l\hat a_k \; ,
\label{ham}
\end{equation}
where $v_{ijkl}$ are the matrix elements of the nucleon-nucleon 
interaction and the indices $i,j,k$ and $l$ label the single-particle states in
some complete orthonormal basis.

An optimal approximation to the ground state for this Hamiltonian is
attained for the wave function $\Phi$ whose energy expectation value
is minimal, that is
\begin{equation}
\delta \langle \Phi | \hat H | \Phi \rangle =0 \;.
\label{varpr}
\end{equation}
The choice that leads to the Hartree-Fock approximation is a wave function 
with the form of a single Slater Determinant,
\begin{equation}
|\Psi\rangle = \prod_{i=1}^A \hat a_i^+ |0\rangle \;,
\label{slat}
\end{equation}  
where the index $i$ refers to a set of single particle states $\phi_i(\vec r)$
to be determined from the variational principle (\ref{varpr}).

If we apply the variational principle (\ref{varpr}) to the Hamiltonian (\ref{ham})
and consider a trial wave function of the form (\ref{slat}), we get a single
particle Hamiltonian, the so called Hartree-Fock Hamiltonian \cite{grei}:
\begin{equation}
h_{kl} = t_{kl} + \sum_{j=1}^A \bar v_{kjlj} = \epsilon_k \delta_{kl} \;.
\end{equation}
By solving these equations, we get the most convenient single particle states. The
Hartree-Fock approximation is often called the self-consistent 
mean-field approximation.

The first feasible Hartree-Fock calculations in nuclei were developed in the 
seventies by Vautherin and Brink when they included the Skyrme interactions 
in the formalism \cite{vaut}. The effective nucleon-nucleon interaction
proposed by Skyrme has two parts: a two-body interaction and a three-body
interaction,
\begin{equation}
V=\sum_{i<j}v_{ij} + \sum_{i<j<k} v_{ijk}\;.
\end{equation}
Skyrme proposed a short-range expansion for $v_{ij}$, that leads to a local
potential depending on the velocity; on the other hand, 
$v_{ijk}$ is equivalent 
to a 2-body density dependent interaction. To obtain the Hartree-Fock 
equations, we have to evaluate the expected value of the Hamiltonian
in a Slater determinant, $E=\langle  HF|\hat H |HF \rangle$, which is
equivalent to computing the spatial integral of a hamiltonian density,
$\int d^3 r \: \mathcal{H}(r) \nonumber$. This hamiltonian density is a
functional of certain densities, $\rho_q(r)$, $\tau_q(r)$, $J_q(r)$,
and depends also on the Skyrme interaction parameters.

In order to take into account the BCS pairing correlations, we consider
a Hamiltonian that contains a pure single-particle part plus a residual 
interaction acting only on Cooper pairs and whose matrix elements are assumed
to be constant:
\begin{equation}
\hat H = \sum_k \epsilon_k^0 \hat a_k^+ \hat a_k - G
\sum_{kk^\prime >0} \hat a_k^+ \hat a_{-k}^+ \hat a _{-k^\prime} \hat
a_{k^\prime}\;.
\end{equation}
In this case there is an approximate solution based on the BCS ground state,
\begin{equation}
|BCS\rangle = \prod_{k>0}
(u_k + v_k \hat a_k^+ \hat a_{-k}^+) |0\rangle \;,
\end{equation}
in which every pair of single-particle levels ($k,-k$) is occupied with
a probability $v_k^2$ and remains empty with probability $u_k^2$. Because
of this, from now on we will talk about quasiparticles instead of particles. 
If we consider the variational principle with the constraint that the
expected value of the number of particles is conserved, 
\begin{equation}
\delta \langle BCS | \hat H - \lambda \hat N |BCS 
\rangle =0 \; ,
\end{equation}
we can obtain the occupation probabilities as well as the quasiparticle
energies. We also get the so-called gap equation,
\begin{equation}
\Delta = \frac{G}{2} \sum_{k>0} \frac{\Delta}{\sqrt{
(\epsilon_k-\lambda)^2 + \Delta^2}}\;,
\end{equation}
that links the strength
of the pairing interaction $G$ to the gap $\Delta$, that can be easily
computed from the experimental masses.
 
By means of the QRPA we will study the excited states of the nucleus.
In Hartree-Fock models the excited states are the
particle-hole excitations, or, in our case, the two-quasiparticle
excitations.
%, with an energy corresponding to the difference between 
%the energy of the last filled state and that of the first empty one.
We consider the variational principle $\delta \langle \Psi | \hat H | \Psi
\rangle = 0$ with the Hamiltonian
\begin{equation}
H=E_0 +\sum_\lambda E_\lambda \alpha_\lambda^+ \alpha_\lambda +
\frac{1}{4} \sum_{\mu \lambda \mu^\prime \lambda^\prime} V_{\mu \lambda
\mu^\prime \lambda^\prime} \alpha_\mu^+ \alpha_\lambda^+ 
\alpha_{\lambda^\prime} \alpha_{\mu^\prime} 
\end{equation}
that admits two-quasiparticle admixtures, 
the $\alpha$'s being the quasiparticle
operators. Our new ground state, $|QRPA\rangle$,
differs from the Hartree-Fock ground state since it already contains
ground state correlations. It is the vacuum of the operator 
$\Gamma_\lambda = \sum_{pn} \left[ X_{pn}^{*\lambda} \alpha_n \alpha_p
- Y_{pn}^{*\lambda} \alpha_n^+ \alpha_p^+ \right]$.

By using this framework, we will describe charge-exchange processes,
as for instance $\beta$-decay, on deformed nuclei. 
It is important to remark that in our formalism 
the quadrupole deformation is not an input parameter, but it is
obtained in a self-consistent way from the mean field. Our single
particle states are expanded in eigenstates of an axially symmetric
harmonic oscillator, as it is explained in \cite{sarr}.

For the kind of processes we are interested in, the
relevant residual interactions are isospin forces giving rise to the
allowed Fermi transitions,
\begin{equation}
V_F(12) = \chi_F (t_1^+ t_2^- + t_1^- t_2^+) \;,
\end{equation}
and spin-isospin ones leading to allowed Gamow-Teller transitions,
\begin{equation}
V_{GT} (12) = \chi_{GT} \sigma_1 \cdot \sigma_2 (t_1^+ t_2^- + t_1^- t_2^+) \;.
\end{equation}
In the QRPA calculation we consider both separable particle-hole 
and particle-particle interactions. 
For the particle-hole part we consider a residual interaction, 
as in \cite{sarr,bert}, and the particle-particle part is a
proton-neutron pairing in the $J^\pi = 1^+$ channel.

Now it is easy to compute Fermi and Gamow-Teller strength distributions for 
each possible final state:
\begin{equation}
B^{F^{\pm}}(\omega) = \sum_f \delta(\omega-\omega_0) |\langle f | 
T_{\pm} |0 \rangle |^2 \;,
\label{fermi}
\end{equation}
\begin{equation}
B^{GT^{\pm}}(\omega) = \left\{ \sum_{\omega_0} \delta(\omega-\omega_0)
|\langle \omega_0| \sigma_0 t^\pm | 0\rangle |^2 +
2 \sum_{\omega_1} \delta(\omega-\omega_1)
|\langle \omega_1| \sigma_1 t^\pm | 0\rangle |^2 \right \} \;.
\end{equation}

\section{Gamow-Teller transitions and nuclear deformation}

\begin{figure}[b]
\centerline{\epsfig{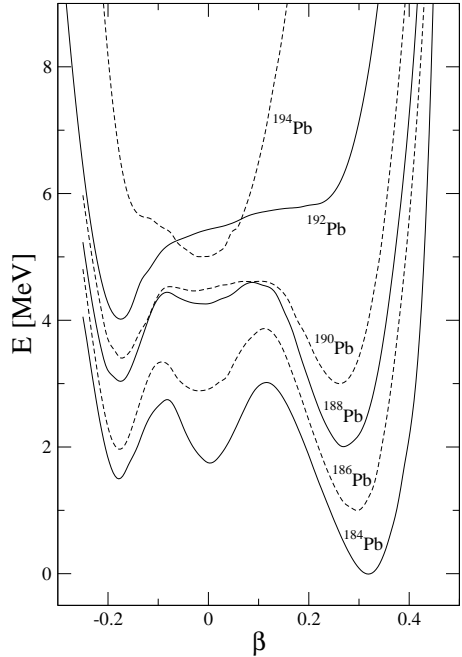}}
\caption{HF energy as a function of the quadrupole deformation $\beta$
for the isotopes $^{184,186,188,190,192,194}$Pb obtained 
from constrained HF+BCS calculations with the force Sk3 and fixed pairing gap
parameters as a function of the quadrupole deformation $\beta$.}
\label{evsq}
\end{figure}

By means of Gamow-Teller strength distributions,
we have studied the signatures of nuclear deformation in $\beta$-decay
patterns in some neutron-deficient Pb, Po and Hg isotopes \cite{more}. 
In fig. \ref{evsq} it is shown
the energy as a function of quadrupole deformation $\beta$ for several
lead isotopes. One can observe that in most cases there are three
minima close in energy: one corresponding to an oblate shape ($\beta < 0$), another
corresponding to a spherical shape ($\beta \approx 0$), and the third one
corresponding to a prolate shape ($\beta > 0$).

\begin{figure}
\centerline{\epsfig{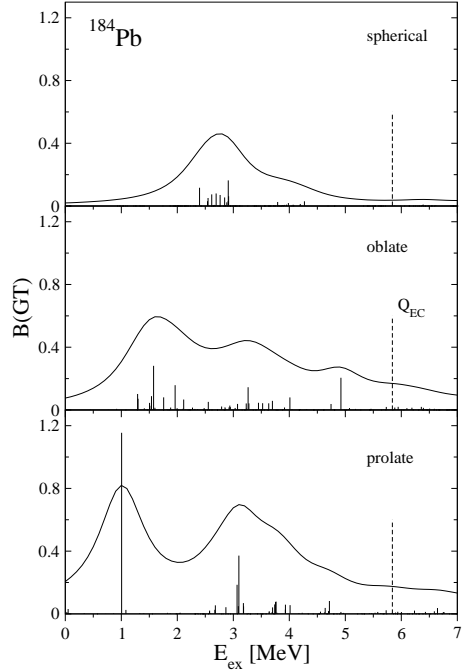}}
\caption{Gamow-Teller strength distributions (discrete and folded) in 
$^{184}$Pb. The experimental $Q_{EC}$ energy is shown with a dashed 
vertical line. 
\label{bgts}}
\end{figure}

We have computed the Gamow-Teller strength distributions for each 
equilibrium deformation and observed that they present 
different features for each equilibrium deformation. 
We observe that the strength is distributed differently
for the various shapes. The distribution is very fragmented in
the oblate case, while it presents a strong single peak at low 
energy in the prolate case. The spherical case shows
a single peak at higher energies. These features can be seen
in fig. \ref{bgts} for the isotope $^{184}$Pb. 
These specific
signatures prevail against changes in Skyrme forces or pairing
treatment and could be used in turn to identify the shape
of a $\beta$ decaying nucleus. Currently the GT strength distributions
for some of these isotopes are
planned to be measured at ISOLDE/CERN \cite{priv} in 2007.

\section{Fermi transitions and isospin mixing}

We have studied Fermi transitions and isospin mixing in some even
Kr isotopes around $N=Z$. We know that the self-consistent mean field
Hamiltonian breaks the symmetries of the exact Hamiltonian, for instance
the total angular momentum is not an exact quantum number for a deformed
nucleus. On the other hand, rotational invariance in isospin space is not an
exact quantum symmetry of the actual total nuclear Hamiltonian, since
the Coulomb force breaks it. We were interested in investigating
how important is this isospin breaking and how
Coulomb interaction and BCS and QRPA correlations contribute to
this mixing. We consider Fermi transitions because for them the effect of 
isosping mixing is more important.

We have performed QRPA calculations on top of a quasiparticle basis 
obtained from a self-consistent deformed Hartree-Fock approach with density
dependent Skyrme interactions, as it has been already explained in section
\ref{frame}. The Fermi strength distribution can be written as in eq.
(\ref{fermi}).

Let us assume now that our ground state $|0\rangle$ is an isospin eigenstate.
The eigenvalues of the isospin operator acting on this state 
must be $T_z=(N-Z)/2$ and $T=\left| T_z\right|$. Hence, it is clear that:
\begin{eqnarray}
T_+ \left| 0\right\rangle &=& T_+ \left| T\, T_z \right\rangle
=\sqrt{(Z-N)}\, \left| T\, T_z +1 \right\rangle \text{ for } N<Z \, ,
\label{t+}   \\
&=&0 \quad \text{ for } N\ge Z \, . \nonumber
\end{eqnarray}
\begin{eqnarray}
T_-\left| 0\right\rangle &=&T_-\left| T\, T_z \right\rangle
=\sqrt{(N-Z)}\left| T\, T_z -1 \right\rangle \text{ for } N>Z \, ,
\label{t-}  \\
&=&0 \quad \text{ for }N\le Z  \, .\nonumber
\end{eqnarray}
In other words, by the $\beta^+$ operator there are no accessible states
if $N \ge Z$ and by the $\beta^-$ operator there are no accessible states
if $Z \ge N$. We will use this ``isospin-forbidden'' transitions in order
to quantify the isospin mixing in our ground state.

The amount of angular momentum mixing \cite{moya} in
the mean field ground state of axially symmetric deformed nuclei is measured
by the expectation value of the squared angular momentum operator
perpendicular to the symmetry axis z,
\begin{equation}
\left\langle J_{\perp }^2\right\rangle \equiv \left\langle
J^2\right\rangle -\left\langle J_z\right\rangle ^2=\frac{1}{2}
\sum_f\left| \left\langle f\right| J_+\left| 0\right\rangle \right|
^2+\left| \left\langle f\right| J_-\left| 0\right\rangle \right| ^2 \, .
\end{equation}
Similarly, taking the z axis in isospin space in the standard way 
($\hat{T}_z=(\hat{N}-\hat{Z})/2$, 
or $T_z=(N-Z)/2$), we can measure the amount of isospin mixing by the
expectation value of  $T_{\perp }^2=T_x^2+T_y^2 = \frac{1}{2}\sum_f\left(\left|
\left\langle f\left| T_+\right| 0\right\rangle
\right| ^2+\left| \left\langle f\left| T_-\right| 0\right\rangle
\right| ^2\right)$, or better
\begin{equation}
\left\langle T_{\perp }^2\right\rangle _0=
\left\langle T_{\perp}^2 \right\rangle -\left| \frac{N-Z}{2}\right| \, .
\end{equation}
In the limit in which the ground state is an isospin eigenstate, 
$T=|T_z|=|(N-Z)/2|$, and $\left\langle T_{\perp }^2\right\rangle _0=0$.
Therefore $\left\langle T_{\perp }^2\right\rangle _0$ is a good way to
measure how important is the isospin mixing.

It can be seen \cite{alva} that the value of $\left\langle T_{\perp }^2
\right\rangle _0$ is purely due to the correlation terms
\begin{equation}
\left\langle T_{\perp }^2\right\rangle _0=\left\langle T_{\perp}^2
\right\rangle -\left| \frac{N-Z}{2}\right| =C_{BCS}+C_{QRPA}\, ,
\label{tfin}
\end{equation}
and that for $N=Z$ and no Coulomb force $\left\langle T_{\perp }^2
\right\rangle _0 \approx 0$.

\begin{table}
\caption{\label{tabl} Amount of isospin mixing for various
approaches in several Kr isotopes.}\smallskip
\begin{center}
\begin{tabular}{cccccc}
 & \multicolumn{2}{c}{\small $\left\langle
T_{\perp }^2\right\rangle _0$ ($\Delta =$ 0.1 MeV)} && 
\multicolumn{2}{c}{\small $\left\langle
T_{\perp }^2\right\rangle _0$ ($\Delta \sim$ 1.5 MeV) } \\ 
\cline{2-3} \cline{5-6} 
& Coul. & No Coul. && Coul. & No Coul.  \\ \hline
$^{70}$Kr   & 0.05 & 0.00 && 0.43 & 0.41 \\ 
$^{72}$Kr   & 0.07 & 0.02 && 1.15 & 1.11 \\ 
$^{74}$Kr   & 0.03 & 0.00 && 0.50 & 0.44 \\ 
%            & 0.05 & 0.00 && 0.46 & 0.48 \\ 
$^{76}$Kr   & 0.02 & 0.00 && 0.30 & 0.23 \\ 
$^{78}$Kr   & 0.01 & 0.00 && 0.17 & 0.12 \\ 
\end{tabular}
\end{center}
\end{table}

\begin{figure}[b]
\centering
\includegraphics[width=3cm,angle=0]{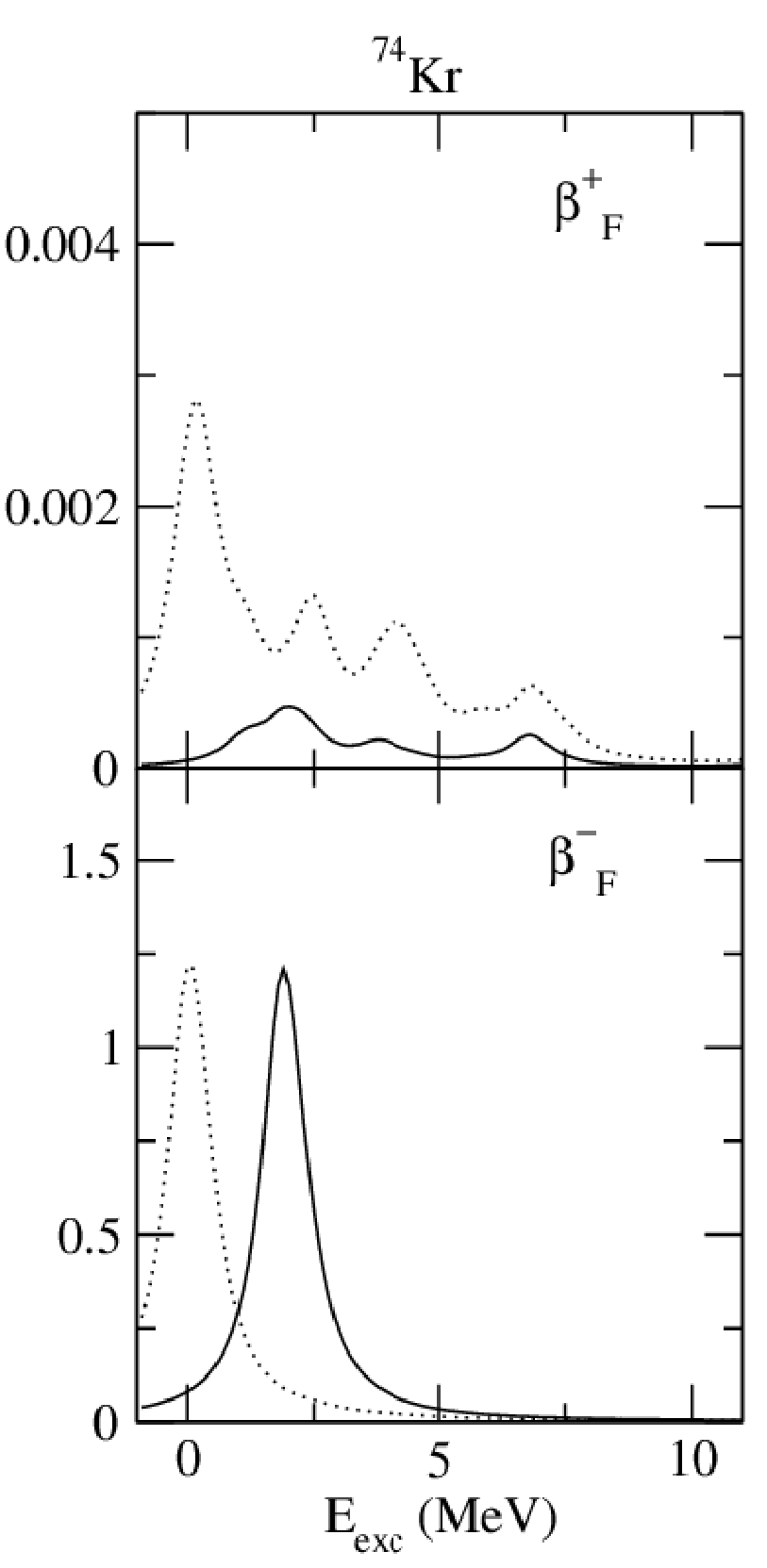}
\hspace{0.1cm}
\includegraphics[width=3cm,angle=0]{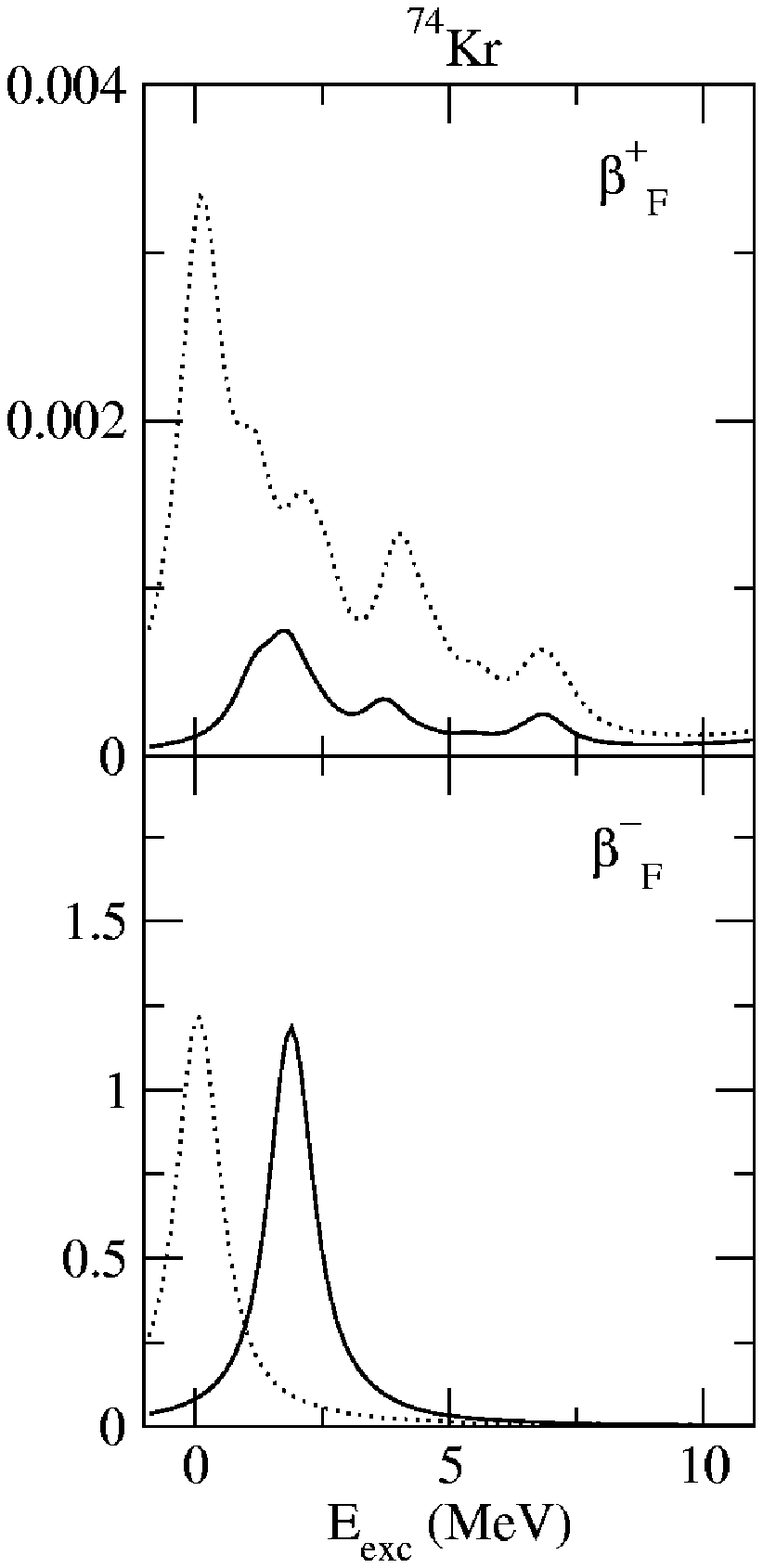}
\hspace{0.1cm}
\includegraphics[width=3cm,angle=0]{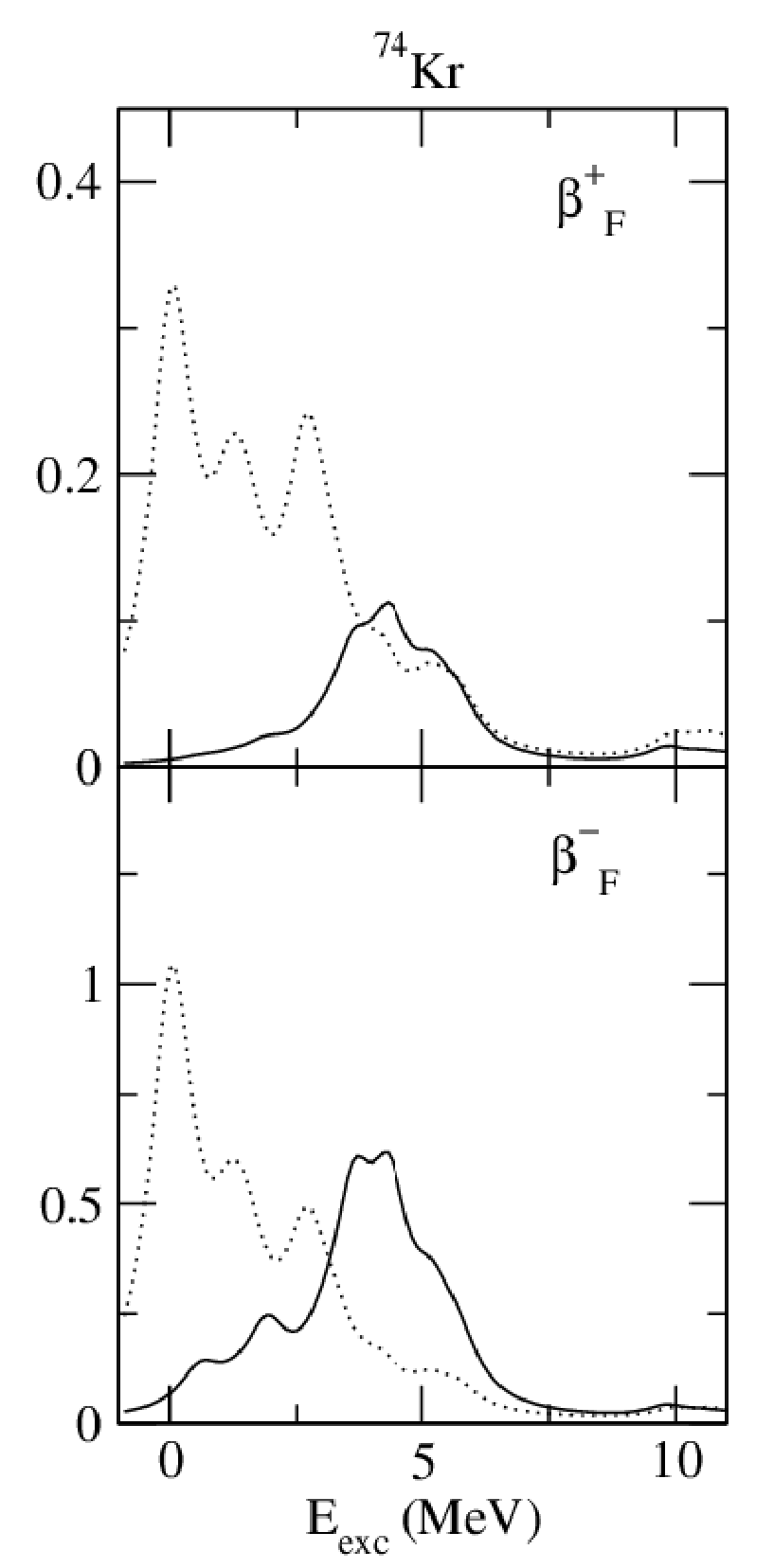}
\caption{Fermi strength distributions plotted as a function of the
excitation energy of the daughter nucleus. The left panels correspond
to the case of no Coulomb interaction and pairing gaps approaching zero.
The middle panels include also Coulomb interaction. The right panels
include both Coulomb interaction and pairing correlations with
realistic gaps.}
\label{figfermi}
\end{figure}

Taking as a reference the case of a self-consistent mean field calculation, 
we can see in table \ref{tabl} that in the absence of Coulomb 
interactions (and in the limit 
of small pairing correlations) the isospin-forbidden transitions 
($\beta ^-$ in $N\le Z$ and $\beta ^+$ in $N\ge Z$) are 
negligible. When the isospin-breaking Coulomb interaction is switched on, 
there is an increase of isospin-forbidden Fermi transitions. Although this 
increase is small, it is a signature of isospin-breaking. Pairing 
correlations increase by orders of magnitude the isospin-forbidden Fermi 
transitions, a fact that is related to the isospin-breaking nature of 
the quasiparticle mean field, which increases with increasing pairing 
gaps. On the other hand, the isospin-breaking effects and forbidden Fermi
transitions are reduced when RPA correlations are taken into account 
(see fig. \ref{figfermi}).

\section{Conclusions}

We have studied Gamow-Teller strength distributions in a search for
signatures of deformation in neutron-deficient lead isotopes, and 
Fermi transitions as a measure of isospin mixing in Kr isotopes.

The theoretical framework is a deformed pnQRPA formalism with spin-isospin
(GT transitions) or isospin-isospin (F transitions) ph and pp separable
residual interactions. The quasiparticle mean field is of Skyrme-HF type
including pairing correlations in BCS approximation. 

Gamow-Teller strength distributions of neutron-deficient lead
isotopes show a strong dependence on nuclear deformation, 
which remains against 
changes of the Skyrme and pairing forces. We conclude that $\beta$-decay
of these isotopes could be a useful tool to look for fingerprints of nuclear
deformation.

We have studied isospin mixing properties in several Kr isotopes 
around $N=Z$ and  analyzed their Fermi transitions at various levels 
of approximation. We have considered 
self-consistent deformed Skyrme HF mean 
fields with and without Coulomb and pairing interactions. Then, we 
take into account isospin-dependent residual interactions and consider 
QRPA correlated ground states. 
We have studied the effect on the Fermi strength distributions of 
isospin-breaking interactions (Coulomb force and pairing), as well as 
the effect of QRPA correlations including Fermi type residual interactions,
whose particle-hole strengths are consistently fixed with the Skyrme force.
Starting from strict Hartree-Fock approach without 
Coulomb, it is shown that the isospin breaking is negligible, on the 
order of a few per thousand for $(N-Z)=6$, increasing to a few percent 
with Coulomb. Pairing correlations induce rather large isospin mixing 
and Fermi transitions of the forbidden type ($\beta ^-$ for $N\le Z$, 
and $\beta ^+$ for $N\ge Z$). The enhancement produced by BCS 
correlations is compensated to a large extent by QRPA correlations 
induced by isospin-conserving residual interactions that tend to 
restore isospin symmetry.

\section*{Acknowledgments}

This work has been supported by Ministerio de Educaci\'{o}n y Ciencia (Spain)
under contract FIS2005-00640. R.A.R. thanks I3P Programme (CSIC, Spain) and 
O.M. thanks Ministerio de Educaci\'on y Ciencia for financial support.

\end{document}